# CMRxRecon2024: A Multi-Modality, Multi-View K-Space Dataset Boosting Universal Machine Learning for Accelerated Cardiac MRI

Zi Wang[1,2,#], Fanwen Wang[2,3,#], Chen Qin[4], Jun Lyu[5], Cheng Ouyang[6], Shuo Wang[7], Yan Li[8], Mengyao Yu[9], Haoyu Zhang[1], Kunyuan Guo[1], Zhang Shi[10], Qirong Li[9], Ziqiang Xu[11], Yajing Zhang[12], Hao Li[13], Sha Hua[14], Binghua Chen[15], Longyu Sun[9], Mengting Sun[9], Qin Li[9], Ying-Hua Chu[16], Wenjia Bai[6], Jing Qin[17], Xiahai Zhuang[18], Claudia Prieto[19,20,21], Alistair Young[20], Michael Markl[22], He Wang[13], Lianming Wu[15], Guang Yang[2,3,20], Xiaobo Qu[23,1,†], Chengyan Wang[24,25,†,*]

[1]Department of Electronic Science, Fujian Provincial Key Laboratory of Plasma and Magnetic Resonance, National Institute for Data Science in Health and Medicine, Xiamen University, Xiamen, China
[2]Department of Bioengineering and Imperial-X, Imperial College London, London, United Kingdom
[3]Cardiovascular Research Centre, Royal Brompton Hospital, London, United Kingdom
[4]Department of Electrical and Electronic Engineering & Imperial-X, Imperial College London, London, United Kingdom
[5]Psychiatry Neuroimaging Laboratory, Brigham and Women's Hospital, Harvard Medical School, Boston, United States
[6]Department of Computing & Department of Brain Sciences, Imperial College London, London, United Kingdom
[7]Digital Medical Research Center, School of Basic Medical Sciences, Fudan University, Shanghai, China
[8]Department of Radiology, Ruijin Hospital, Shanghai Jiao Tong University School of Medicine, Shanghai, China
[9]Human Phenome Institute, Fudan University, Shanghai, China
[10]Department of Radiology, Zhongshan Hospital, Fudan University, Shanghai, China
[11]School of Health Science and Engineering, University of Shanghai for Science and Technology, Shanghai, China
[12]GE Healthcare, Beijing, China
[13]Institute of Science and Technology for Brain-Inspired Intelligence, Fudan University, Shanghai, China
[14]Department of Cardiovascular Medicine, Ruijin Hospital Lu Wan Branch, Shanghai Jiao Tong University School of Medicine, Shanghai, China
[15]Department of Radiology, Ren Ji Hospital, School of Medicine, Shanghai Jiao Tong University, Shanghai, China
[16]Siemens Healthineers Ltd., Shanghai, China
[17]School of Nursing, The Hong Kong Polytechnic University, Hong Kong, China
[18]School of Data Science, Fudan University, Shanghai, China
[19]School of Engineering, Pontificia Universidad Católica de Chile, Santiago, Chile

[20]School of Biomedical Engineering and Imaging Sciences, King's College London, London, United Kingdom

[21]Millenium Institute for Intelligent Healthcare Engineering, Santiago, Chile

[22]Department of Radiology, Feinberg School of Medicine, Northwestern University, Chicago, United States

[23]Department of Radiology, the First Affiliated Hospital of Xiamen University, School of Medicine, Xiamen University, Xiamen, China

[24]Human Phenome Institute and Shanghai Pudong Hospital, Fudan University, Shanghai, China

[25]International Human Phenome Institute (Shanghai), Shanghai, China

[#] **These authors contributed equally to this work.**

[†] **These authors are co-senior author.**

[*]**Correspondence to:**

Chengyan Wang, Ph.D.

Fudan University

Address: 825 Zhangheng Road, Pudong New District, Shanghai, 201203

Phone: 86-021-3420-8697

E-mail: wangcy@fudan.edu.cn

ORCID: 0000-0002-8890-4973

# CMRxRecon2024: A Multi-Modality, Multi-View K-Space Dataset Boosting Universal Machine Learning for Accelerated Cardiac MRI

**Summary**

The released CMRxRecon2024 dataset is currently the largest and most protocol-diverse publicly available k-space dataset including multi-modality and multi-view cardiac MRI data from 330 healthy volunteers, and each one covers standardized and commonly used clinical protocols.

**Key Points**

1) The CMRxRecon2024 dataset is the largest and most protocol-diverse publicly available cardiac k-space dataset to date, created to facilitate the technical development, fair evaluation, and clinical transfer of cardiac MRI reconstruction approaches.

2) The dataset includes k-space and images from 330 healthy volunteers, covering commonly used modalities, anatomical views, and acquisition trajectories in clinical cardiac MRI workflows.

3) An open platform with tutorials, benchmarks, and data processing tools is provided to facilitate data usage, advanced method development, and fair performance evaluation.

**Abbreviations**

LAX = long-axis, SAX = short-axis, LVOT = left ventricle outflow tract, ECG = electrocardiogram, TrueFISP = true fast imaging with steady state precession, MOLLI = modified Look-Locker inversion recovery, FLASH = fast low angle shot.

# 1. Introduction

MRI is currently the reference standard imaging modality for non-invasive and non-radioactive cardiovascular disease diagnosis. Cardiac MRI has emerged as a clinically crucial technique for evaluating cardiac morphology, function, perfusion, viability, and quantitative myocardial tissue characterization owing to its ability to provide diverse information with multiple modalities and detailed anatomical views (1-5). Despite these advantages, cardiac MRI suffers from prolonged data acquisitions due to the need for high spatiotemporal resolution, high-dimension, various modalities, and extensive whole-heart coverage, compounded with the physical limitation of imaging systems. Accelerating cardiac MRI facilitates the achievement of high spatiotemporal resolution, improvement of patient comfort, and reduction of motion-induced artifacts. Advanced image reconstruction approaches are essential to recover high-quality, clinically interpretable images from highly undersampled k-space data (3, 6-8).

Recently, artificial intelligence techniques, particularly deep learning, have shown great potential in cardiac MRI reconstruction (9-14), but the development of these techniques are currently limited by the lack of large-scale publicly available datasets that contain raw k-space measurements (15). Although several public cardiac MRI datasets contain k-space for benchmarking (16-18) and have facilitated deep learning approaches (14), they are limited to only three modalities (i.e., cine, T1/T2 mapping) with restricted views, highlighting the insufficient diversity and quantity of such data. To date, most cardiac MRI reconstruction models are trained and validated on carefully pre-processed datasets with specific imaging scenarios. As a result, these models struggle to handle the diverse and complex scenarios in clinical practice, hindering substantial technological progress and limiting their widespread applications.

The purpose of the CMRxRecon2024 dataset is to move toward addressing data accessibility and diversity issues in cardiac MRI reconstruction. To this end, we build a protocol-diverse cardiac MRI dataset, involving multiple modalities, anatomical views, and k-space undersampling trajectories to promote the clinical translation of image reconstruction approaches. The dataset covers multiple cardiac imaging scenarios, facilitating the evaluation of the generalization performance of emerging frameworks, and providing data support for future universal model developments. In addition, to facilitate the use of released dataset and to promote fair performance evaluation, an open platform with tutorials, benchmarks, and data processing tools is provided.

**2. Materials and Methods**

Here, we describe our recently released protocol-diverse dataset tailored for multi-scenario cardiac MRI reconstruction (Figure 1). Our dataset includes raw multi-coil MRI k-space data from 330 healthy volunteers. Each one has multi-modality k-space data consisting of cardiac cine, T1/T2 mapping, tagging, phase-contrast (i.e., flow2d), and black-blood imaging, covering commonly used clinical protocols. It also includes different anatomical views like long-axis (LAX: 2-chamber, 3-chamber, and 4-chamber), short-axis (SAX), left ventricle outflow tract (LVOT), and aorta (transversal and sagittal views). Notably, various k-space undersampling trajectories (i.e., uniform, Gaussian, and pseudo radial) with different acceleration factors are provided for retrospective undersampling. Figure 2 shows the overall workflow to prepare our CMRxRecon2024 dataset, from data acquisition to the final released dataset.

## 2.1 Data Acquisition

The study received approval from our local institutional review board (approval number: MS-R23). As part of the written consent process, participants agreed to make their anonymized data publicly available. All participants were informed about the study's nature and consented to share their materials in anonymized form. The enrollment process and screening protocols were as follows: 1) Enrollment: All volunteers were recruited from the local community, rather than through hospitals, meaning they were not patients but individuals volunteering for scientific research. The inclusion criteria were: i) adults over 20 years old without a pathologically confirmed diagnosis of cardiovascular disease, and ii) availability of an MRI examination with all necessary imaging sequences. 2) Screening: Scans with incomplete data and/or identified by the technicians to have substantial artifacts were excluded. Between June 2023 and February 2024, 330 healthy Asian volunteers provided written informed consent and participated in the study.

The imaging data were prospectively acquired for this research with specifically designed multi-modality and multi-view protocols. Data were acquired using a 3T scanner (MAGNETOM Vida, Siemens Healthineers), equipped with dedicated multi-channel cardiac coils (19, 20). Participants were positioned supine on the table before the scans. Electrodes were attached, and electrocardiogram (ECG) signals were recorded during the scanning process. Sedation was not required during the acquisition process for any of the participants. The 'Dot' engine was utilized for cardiac scout imaging. We adhered to the cardiac MRI recommendations outlined in the previous publications (18, 21). As shown in Figure 3, data were acquired with six modalities containing different anatomical views: (a) cine imaging with seven anatomical views, namely LAX (2-chamber, 3-chamber, and 4-chamber), SAX, LVOT, and aorta (transversal and sagittal views), (b) phase-contrast (i.e., flow2d) with transversal view, (c) tagging with SAX view, (d) black-blood with SAX view, (e) T1 mapping with SAX view, and (f) T2 mapping with SAX view.

The typical acquisition parameters of imaging protocols are summarized in Table 1. (a) The TrueFISP sequence was used for cine, phase-contrast (i.e., flow2d), and tagging acquisitions under breath-hold. They were acquired through a retrospective ECG-gated segmented approach, wherein k-space was segmented in the phase encoding direction across multiple cardiac cycles. The selection of breath holds was automatically optimized according to the acquisition size, slice, and heart rate. (b) The modified Look-Locker inversion recovery-fast low angle shot (MOLLI-FLASH) sequence was used for T1 mapping under breath-hold. The 4-(1)-3-(1)-2 scheme with one heart-beat rest was used to obtain nine images with different T1 weightings at the end of the cardiac diastole with ECG triggering. The inversion time varied among participants according to the real-time heart rate. (c) The T2-prepared (T2prep)-FLASH sequence was used for T2 mapping under breath-hold. Three images with different T2 weightings were acquired at the end of the cardiac diastole with ECG triggering. T2 preparation time was 0/35/55 ms. (d) The turbo spin echo (TSE) sequence was used for black-blood under breath-hold. The image with blood flow suppression was acquired at the end of the cardiac diastole with ECG triggering.

*2.2 Data Preparation*

Here, we briefly introduce the general workflow to produce our CMRxRecon2024 dataset from the scanner. Specifically, the raw data with the filename extension '.dat' was exported from the scanner using the Siemens software TWIX directly. The k-space data were then extracted using the Matlab toolbox mapVBVD (https://github.com/pehses/mapVBVD). The k-space data were anonymized via conversion to the raw data format. We removed all information related to participant identity, e.g., participant name, hospital location, date of exam and birth. The individual k-space lines are already correctly sorted according to their position in the acquisition trajectory, and no other preprocessing steps were performed. Image quality control was carefully carried out by two radiologists (Y. L. and S. H., with 5 and 6 years of clinical experience) through visual assessment, to remove low-quality images with obvious motion, magnetic susceptibility, metal, and aliasing artifacts. After these processing steps, the resulting k-space was transformed to the '.mat' Matlab format.

Table 2 offers an overview of the key metadata fields ('csv' format) provided with the k-space data, including acquisition hardware, acquisition k-space, and sequence parameters. We also released a Github repository (https://github.com/CmrxRecon/CMRxRecon2024) that provides tools to load and reconstruct k-space data, using the commonly used programming languages (i.e., Matlab and Python). Since the data were acquired using multi-channel receiving array coils, correctly combining the images from each coil is a crucial step in the image reconstruction (19, 20). An additional calibration step was required to obtain coil sensitivity information. To avoid bias towards specific methods for estimating coil sensitivity maps and to control the overall dataset size, the coil sensitivity maps were not included in our dataset. However, we provided a typical example of using ESPIRiT (22) for coil sensitivity estimation in our Github repository, allowing researchers from different communities to quickly get started.

In our released dataset for open evaluation, the k-space data of 330 healthy volunteers were partitioned into the following three components: (a) training dataset with 200 individuals. (b) validation dataset with 60 individuals, and (c) test dataset with 70 individuals. The training dataset can be used to train reconstruction models and to determine hyperparameters, while the validation and test datasets can be used to compare the results across different approaches. Open evaluation on the validation and test datasets were accomplished by uploading reconstruction results to a public leaderboard: https://www.synapse.org/#!Synapse:syn54951257/wiki/627149. Notably, since training, validation, and testing data followed the same processing procedures, researchers can easily use these data for their own studies in any combination.

To simulate different acceleration scenarios, various k-space undersampling trajectories (i.e., uniform, Gaussian, and pseudo radial) with different acceleration factors (i.e., 4~24) were provided for retrospective undersampling (13, 14, 23). The validation and test datasets contained undersampled k-space data. The undersampling was implemented by retrospectively applying masks to fully-sampled multi-coil k-space data, and the acceleration factors were calculated without including central autocalibration signals. Notably, all slices from the same individual were assigned an identical undersampling mask, while different individuals received randomly selected masks to ensure diversity in undersampling trajectories. Figure 3 shows typical undersampling masks. The processes for generating undersampling masks and conducting retrospective undersampling are provided in our Github repository. This resource aims to enable a broader exploration of undersampling scenarios and assist in integrating cardiac MRI into complex clinical workflows.

## 3. Resulting Dataset

The released CMRxRecon2024 dataset is the largest and most protocol-diverse publicly available cardiac k-space dataset to date. Data were acquired from 330 healthy volunteers, with a mean age of 36 ± 12 years and mean body mass index (BMI) of 23.35 ± 3.46 kg/m$^2$. The age range of included individuals was 20-60 years, with 40.3% (133/330) of individuals aged 20-30 years, 24.5% (81/330) aged 30-40 years, 16.1% (53/330) aged 40-50 years, and 19.1% (63/330) aged 50-60 years. In terms of BMI, 3.03% (10/330) of individuals in the dataset were considered underweight, 70.0% (231/330) healthy, 20.3% (67/330) overweight, and 6.67% (22/330) obese. Regarding sex distribution, 47.3% (156/330) of individuals were female and 52.7% (174/330) were male.

The dataset covers commonly used modalities (cardiac cine, T1/T2 mapping, tagging, phase-contrast, and black-blood imaging), anatomical views (long-axis with 2-chamber, 3-chamber, and 4-chamber, short-axis, left ventricle outflow tract, and aorta with transversal and sagittal views), and acquisition trajectories (uniform, Gaussian, and pseudo radial sampling with different acceleration factors) in clinical cardiac MRI workflows. The CMRxRecon2024 dataset can be downloaded at https://www.synapse.org/#!Synapse:syn54951257/wiki/627141.

In addition to serving as a data portal, this Synapse repository can also be used for online performance evaluations and discussion forums.

Moreover, to facilitate data usage, advanced method development, and fair performance evaluation, the tutorials, benchmarks, and data processing tools are provided in the Github repository: https://github.com/CmrxRecon/CMRxRecon2024. The performance evaluation includes calculating three commonly used metrics between the reconstructed images and the reference standard (i.e., fully sampled images): structural similarity index measure (SSIM) (24), peak signal-to-noise ratio (PSNR), and normalized mean squared error (NMSE). A higher SSIM, higher PSNR, and lower NMSE indicate better image detail preservation, less image distortion, and lower reconstruction error, respectively. For the benchmark, two conventional parallel reconstruction algorithms, GRAPPA (20) and SENSE (19), are provided.

The dataset is openly accessible to individuals for educational and research purposes, and registered users can access it without requiring approval. Notably, although the commercial use of the dataset itself is prohibited, we do not restrict the use of the dataset for developing, testing, or refining software, algorithms, or other intellectual property for academic research.

**Discussion**

To the best of our knowledge, our CMRxRecon2024 dataset is the largest and most protocol-diverse publicly available k-space dataset of cardiac MRI, covering six modalities, seven anatomical views, and four types of acquisition trajectories. Previous publicly available cardiac k-space datasets have limited modalities and anatomical views, such as the OCMR dataset (cine with four views) (16), Harvard cardiac dataset (cine with one view) (17), and CMRxRecon dataset (cine and T1/T2 mapping with four views) (18). Notably, our dataset does not overlap with the individuals included in the CMRxRecon dataset (18) released in 2023.

Our goal is to provide a standardized, protocol-diverse, and high-quality dataset to facilitate technical development, fair evaluation, and clinical transfer of cardiac MRI reconstruction approaches. We hope to promote the development and validation of universal image reconstruction frameworks that enable fast and robust reconstructions across diverse cardiac MRI protocols in clinical practice. We would like to highlight that the CMRxUniversalRecon challenge at MICCAI 2024, which is based on our CMRxRecon2024 dataset, has already concluded. The top-5 reconstruction methods trained on this dataset have significantly outperformed the previous benchmarks, GRAPPA (20) and SENSE (19), in three commonly used evaluation metrics (i.e., SSIM, PSNR, and NMSE). See the leaderboard at: https://www.synapse.org/Synapse:syn54951257/wiki/627936. These improvements demonstrate the dataset's usability, while also preliminarily showing its potential to facilitate advancements in learning-based reconstruction methods.

Currently, the CMRxRecon2024 dataset consists of multi-modality cardiac MRI data from healthy Asian volunteers and all data were collected from a single vendor (3T MAGNETOM Vida, Siemens Healthineers) in a single center. Thus, there are three main limitations of our dataset are summarized as follows. First, it includes data from a single-vendor and single-center, which may restrict the generalizability of deep learning models to other MRI systems and lead to performance degradation in multi-vendor and multi-center imaging. Second, only healthy volunteers were included in the dataset. We chose to include only healthy individuals to ensure data throughput and quality, since these individuals generally have a longer tolerance for complex MRI scans and breath-hold. However, this inclusion criterion leads to the inability to represent the full diversity of clinical conditions, such as different cardiovascular diseases, which is crucial for developing models applicable to patient populations. Finally, the dataset lacks ethnic diversity, as it includes only Asian individuals. This homogeneous population may limit the model adaptability to more ethnically diverse populations.

Considering the complexity and diversity of cardiac MRI, there remain many issues for the research community to further explore, which puts higher demands on the available dataset. We are planning to progressively add new data to the repository during future releases. The preliminary plan is to involve multi-vendor and multi-center protocols, typical cardiovascular diseases, and to cover diverse populations and clinical applications. These efforts will improve the representativeness of the dataset and its applicability to a broader range of individuals. Specifically, the next dataset will include data from at least 5 medical centers worldwide, with k-space and images acquired using over 10 different scanners with field strengths of 1.5T, 3.0T, and 5.0T. These scanners will come from 4 mainstream vendors: GE, Philips, Siemens, and United Imaging. We plan to cover at least 5 cardiovascular pathologies, including but not limited to myocardial infarction, cardiomyopathy, and atrial fibrillation. The planned scanning population is 600 cases, with at least 100 cases per center, from Asia, Europe, and the Americas to ensure ethnic diversity.

We believe that the availability of our CMRxRecon2024 dataset will expedite research in multi-modality cardiac MRI reconstruction, in parallel with image reconstructions of brain and knee MRI that are boosted by well-curated, large-scale datasets from the fastMRI-family (25, 26). It can serve as a benchmark for training and evaluating new approaches and as an example and incentive for upcoming public datasets to further address the accuracy and generalizability issues of deep learning in image reconstruction. In summary, the CMRxRecon2024 dataset will substantially aid in accelerating the deployment of advanced models and facilitating clinical adaptation to achieve more time-efficient, patient-friendly, and reliable diagnosis of cardiovascular diseases.

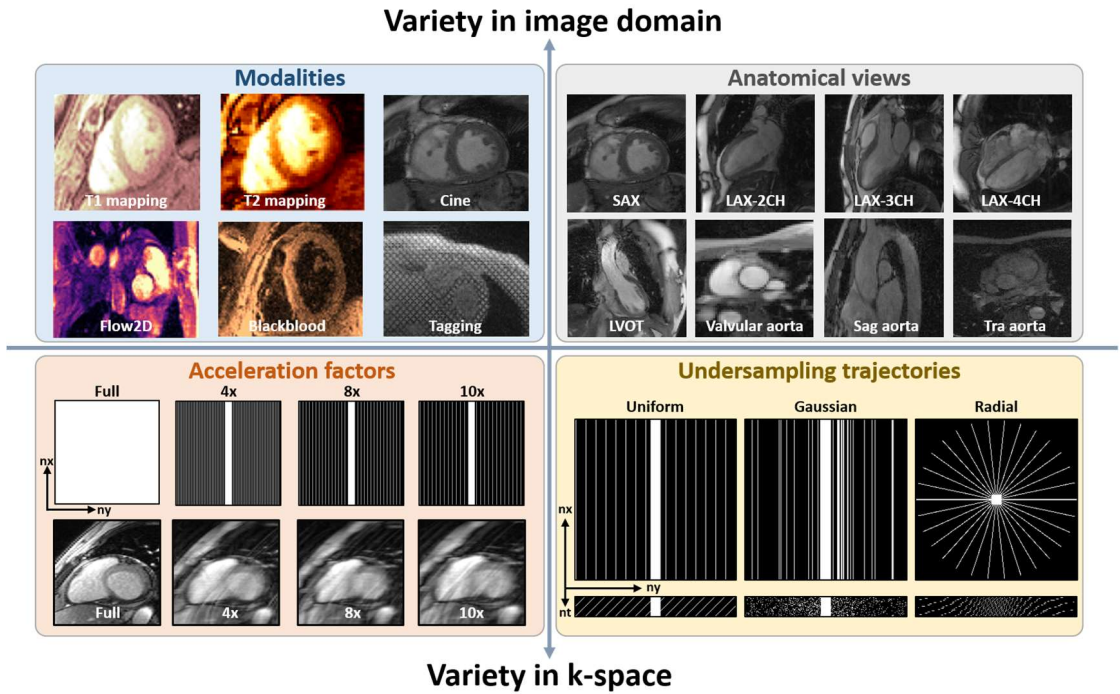

Figure 1: An overview of the released CMRxRecon2024 dataset. The dataset includes multi-modality cardiac MRI with diverse anatomical views. With various under sampling trajectories and acceleration factors, deep learning-based reconstruction methods can be developed for high spatiotemporal resolution images and comprehensive cardiac assessment in reduced scanning time. SAX = short axis, LAX = long axis, 2CH = 2 chamber, 3CH = 3 chamber, 4CH = 4 chamber, LVOT = left ventricle outflow tract, Sag = sagittal view, Tra = transversal view.

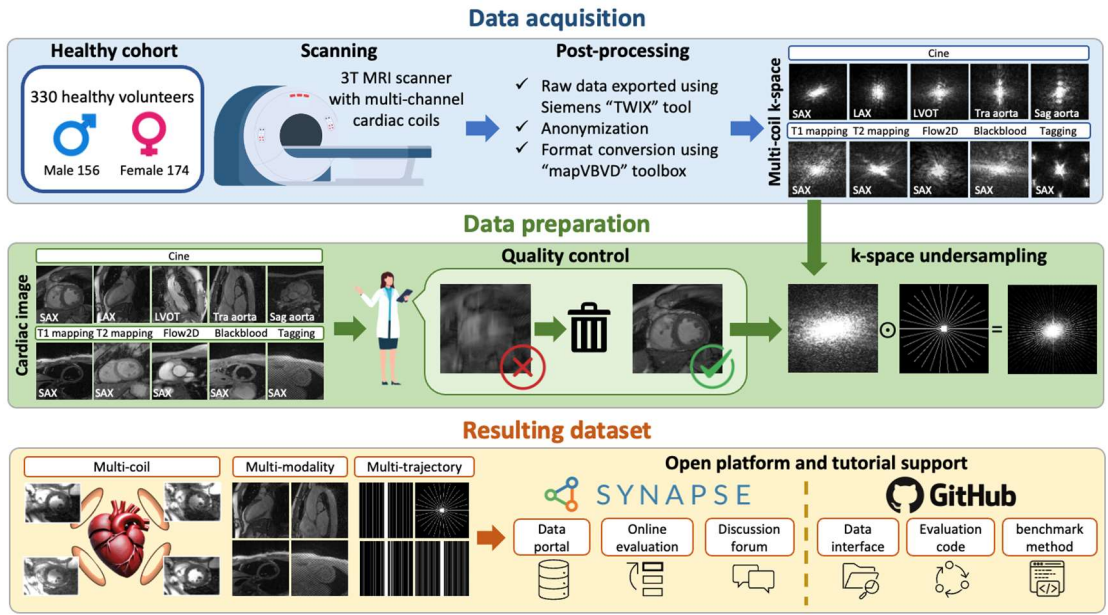

Figure 2: The workflow to prepare the CMRxRecon2024 dataset, from data acquisition to the final released dataset. Multi-coil, multi-modality, and multi-view k-space data were acquired from 330 healthy volunteers using a 3T MRI scanner with multi-channel cardiac coil.

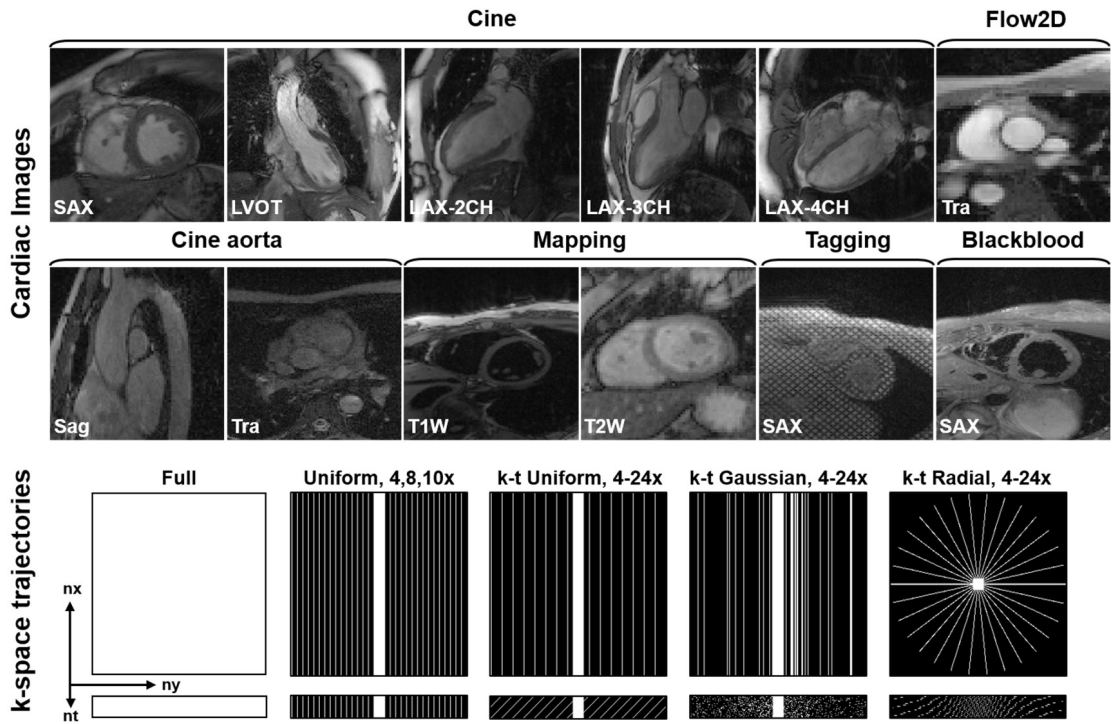

Figure 3: Example images of multi-modality cardiac MRI with various anatomical views and undersampling trajectories. T1W = T1 weighted, T2W = T2 weighted, LAX = long axis, 2CH = 2 chamber, 3CH = 3 chamber, 4CH = 4 chamber, SAX = short axis, LVOT = left ventricle outflow tract, Tra = transversal view, Sag = sagittal view.

**Table 1: Acquisition parameters for the imaging protocols used to acquire k-space data represented in the CMRxRecon2024 dataset.**

|  | Cine LAX | Cine SAX | Cine LVOT | Cine Sag aorta | Cine Tra aorta | Tagging SAX | Flow2D SAX | T1map SAX | T2map SAX | Blackblood SAX |
|---|---|---|---|---|---|---|---|---|---|---|
| Sequence | TrueFISP | TrueFISP | TrueFISP | TrueFISP | TrueFISP | SPAMM-TrueFISP | Venc-TrueFISP | MOLLI-FLASH | T2prep-FLASH | TSE |
| FOV X (mm) | 340-383 | 344-404 | 340 | 300 | 300 | 344 | 360 | 360-380 | 360 | 340 |
| FOV Y (mm) | 236.79-379.58 | 215-404 | 304.04 | 302.88 | 302.88 | 340.93 | 360 | 125 | 288.75-304.79 | 265.63 |
| Acq X | 352 | 404 | 328 | 328 | 328 | 372 | 288 | 404 | 320 | 512 |
| Acq Y | 56-58 | 54-82 | 56 | 56 | 56 | 90 | 72 | 125 | 86 | 78 |
| No. of slices | 3 | 8-14 | 1 | 2-11 | 8-10 | 3-15 | 2 | 1 | 1 | 5-7 |
| Slice thickness (mm) | 6 | 8 | 6 | 3 | 6 | 30-34 | 6 | 5 | 5 | 5 |
| No. of coils | 30 | 30 | 30-34 | 30-34 | 30-34 | 15-41 | 34 | 30 | 30-34 | 30 |
| Temporal phase | 14-42 | 14-36 | 18-55 | 16-48 | 17-45 | 2 | 18-51 | 9 | 3 | 1 |
| TR (ms) | 39.96-43.80 | 45.78-47.88 | 39.24 | 43.08 | 40.44 | 47.61 | 36.64 | 358.40-359.48 | 2 | 577-800 |
| TE (ms) | 1.46-1.57 | 1.44-1.50 | 1.43 | 1.63 | 1.47 | 2.54 | 2.50 | 1.13 | 202.66-207.82 | 44 |
| Flip angle (°) | 39-52 | 37-44 | 42-46 | 37-43 | 36-43 | 10 | 20 | 35 | 1.28-1.35 | 180 |

Note: Because not all parameters are completely identical for the different MRI scanners that were used during data acquisition, a range of sequence parameters is shown in some cases. FOV = field of view, TE = echo time, TR = repetition time, 2D = two dimensional, LAX = long axis, SAX = short axis, LVOT = left ventricle outflow tract, SPAMM = spatial modulation of magnetization, Tra = transversal view, Sag = sagittal view, TSE = turbo spin echo, Venc = velocity encoding, FLASH = fast low angle shot, MOLLI = modified Look-Locker inversion recovery, Acq = acquisition matrix, No. = number.

**Table 2: Overview of selected metadata fields that are provided with the k-space data.**

| Category | Raw data |
|---|---|
| Acquisition hardware | Field strength (T) |
|  | Software version |
|  | No. of receive coils |
| Encoded k-space | Acq X |
|  | Acq Y |
|  | FOV X (mm) |
|  | FOV Y (mm) |
|  | Slice thickness (mm) |
|  | Temporal phase |
|  | Slice number |
| Reconstructed image space | Reconstructed matrix X |
|  | Reconstructed matrix Y |
| Sequence parameters | Repetition time (ms) |
|  | Echo time (ms) |
|  | Flip angle (°) |

Note: Acq = acquisition matrix, No. = number.

**Acknowledgments**

This study was supported in part by the National Natural Science Foundation of China (No. 62331021, 62371413, 62122064), the Shanghai Municipal Science and Technology Major Project (No. 2023SHZD2X02A05), the Shanghai Rising-Star Program (No. 24QA2703300), the Royal Society (No. IEC\NSFC\211235), the UKRI Future Leaders Fellowship (No. MR/V023799/1), the EPSRC UK Grants (No. EP/X039277/1), and the China Scholarship Council (No. 202306310177). The computations in this research were performed using the CFFF platform of Fudan University.

**Author contributions**

Chengyan Wang, Guang Yang, Chen Qin, Shuo Wang, Jun Lyu, He Wang, Xiaobo Qu, Xiahai Zhuang, Wenjia Bai, Jing Qin, Alistair Young, Michael Markl, and Claudia Prieto designed the research; Chengyan Wang, Mengting Sun, Qirong Li, and Meng Liu performed data anonymization; Hao Li, Longyu Sun, and Ying-Hua Chu collected data; Chengyan Wang, Zi Wang, Fanwen Wang, Haoyu Zhang, Kunyuan Guo, and Ziqiang Xu analyzed data; Shuo Wang, Yan Li, Qing Li, Zhang Shi, Yajing Zhang, Lianming Wu, Binghua Chen, and Sha Hua performed quality control of the data; Zi Wang, Fanwen Wang, and Chengyan Wang wrote the paper; all authors revised and corrected the manuscript. Zi Wang and Fanwen Wang contributed equally to the paper, and Chengyan Wang is the corresponding author of this manuscript.

**Competing interests**

None.